# Anomalous Proteinaceous Shells with Octagonal Local Order

Sergei B. Rochal[1,*], Aleksey S. Roshal[1], Olga V. Konevtsova[1] and Rudolf Podgornik[2,3]

[1]Faculty of Physics, Southern Federal University, 344090 Rostov-on-Don, Russia

[2] School of Physical Sciences and Kavli Institute for Theoretical Sciences, University of Chinese Academy of Sciences, 100049 Beijing, China

[3] CAS Key Laboratory of Soft Matter Physics, Institute of Physics, Chinese Academy of Sciences, 100190 Beijing, China

Proteinaceous shells useful for various biomedical applications exhibit a wide range of anomalous structures that are fundamentally different from icosahedral viral capsids described by the Caspar-Klug paradigmatic model. Exploring the Protein Data Bank, we have identified nine different types of anomalous shells structurally close to flat octagonal quasicrystals. As we show, these numerous shells have cubic nets cut from short-period approximants of an octagonal tiling composed of square and rhombic tiles. The approximants and parent tiling are easily obtained within the Landau density wave approach, while the nonequilibrium assembly of them can be simulated using the pair potentials derived from critical density waves. Gluing a polyhedron net and mapping it onto a spherical surface induces tile distortions, and to reduce them, we introduce and minimize the effective elastic energy of the system. Thus, we return quasi-equivalence to previously equivalent tiles. Possible cubic faceting of the octagonal spherical tilings is discussed in terms of the topological charge distribution over the tiling vertices. The proposed structural models describe numerous proteinaceous shells including about half of the known symmetrical enzymes. Our results constitute a fundamental basis for further applications of identified octagonal assemblies and can help to discover and study similar systems in the future.

## I. INTRODUCTION

The most common spherical protein assemblies in nature are viral shells (capsids) that protect and transport the viral genome [1]. Their size ranges from 17 to several thousand nm, and the number of proteins vary from 60 to several thousand [2]. Structures of most of these shells are based on icosahedral triangulations of the sphere; all cells of such triangulations are approximately equal equilateral triangles, with three proteins per cell [3]. Locally, such triangulations possess translational symmetry, and in this sense, viral shells resemble 2D crystals.

In contrast to crystals, whose structures are obtained by periodic repetition of a single cell, structures of quasicrystals are formed by few different cells (named as tiles) arranged in an aperiodic manner [4]. Similar anomalous protein shells are also known, and several of them have a dodecagonal square-triangular local order [5]. Other types of quasicrystalline order are also possible in anomalous shells [6-8]. For example, the arrangement of protein positions in the unique viral family BPV demonstrates the chiral decagonal order [8]. As we show in this work, anomalous protein shells with local octagonal order exist and are quite common. Shells of this type are comparable in size to small icosahedral viruses and are highly symmetrical but instead of icosahedral symmetry they have octahedral or tetrahedral ones. Some of these assemblies are chaperones [9-10], but most act as enzymes with different functions [11-14] or serve as nanocontainers for various practical applications [15-17]. Note that the icosahedral symmetry combined with the local periodicity of viral shells may provide a larger internal volume for genome storage. However, the fact that the number of proteins in an icosahedral shell must be a multiple of 60 is probably incompatible with other functions of the anomalous nanocontainers under consideration.

Recall that the famous Caspar and Klug (CK) theory [3] explains the structure of icosahedral viral shells as a result of mapping the flat periodic hexagonal order onto a sphere via icosahedral nets. The specific symmetry of the flat order ensures smooth gluing of the icosahedron edges after removing 60-degree sectors from the net. However, to transfer the octagonal quasicrystalline order onto the sphere, the aperiodic order must be reconstructed into structurally close periodic approximants, which can be mapped onto the sphere via appropriate cubic nets.

In what follows, developing the approach of critical density waves, we first propose a new thermodynamic model of nonequilibrium assembly of random octagonal quasicrystalline tilings and periodic approximants. Then, using the cubic nets cut from short-period approximants, we model octahedral and tetrahedral spherical packings and demonstrate structural relationships between known protein shells and the approximants. Also, we predict the structures of similar, but not yet discovered shells.

## II. PLANAR OCTAGONAL STRUCTURES: CRITICAL DENSITY WAVES AND SELF-ASSEMBLY

Coordinates of vertices of a simple octagonal tiling can be expressed as

$$\mathbf{r}_j = \sum_{i=0}^{3} n_i^j \mathbf{a}_i^{||}, \qquad (1)$$

where $n_i$ are integers and $\mathbf{a}_i^{||} = \langle \cos \pi i/4, \sin \pi i/4 \rangle$. The perpendicular coordinates $\mathbf{r}_j^\perp$ of the same point

*Contact author: rochal_s@yahoo.fr

are determined by the same set $\{n_i^j\}$, but the vectors $\mathbf{a}_i^{\|}$ are replaced by $\mathbf{a}_i^{\perp}$, where $\mathbf{a}_i^{\perp} = \langle \cos 5i\pi/4, \sin 5i\pi/4 \rangle$ [18]. Note that the perpendicular coordinates introduced in this way assume that there is a 4D cubic lattice whose nodes are indexed by four integer indices, while the coordinates (1) and perpendicular coordinates are the mutually orthogonal projections of these nodes. A node is included in the tiling if its perpendicular coordinates belong to an acceptance domain (AD). In the well-known octagonal Ammann-Beenker tiling (ABT) formed by square and rhombic tiles [19], the AD has the shape of a regular octagon with the distance between the opposite sides equal to $\sqrt{2}+1$. If the octagon center coincides with the origin in the perpendicular space, the tiling is symmetric relative to the origin of the direct space; see Fig. 1(a). Since in a primitive 4D cubic lattice the basis vectors of the reciprocal and direct spaces, $\mathbf{b}_i$ and $\mathbf{a}_i$ are parallel to each other and $\mathbf{b}_i \mathbf{a}_j = \delta_{ij}$, then after the 4D lattice has been projected into two subspaces, the latter equation takes the form $\mathbf{b}_i^{\|} \mathbf{a}_j^{\|} + \mathbf{b}_i^{\perp} \mathbf{a}_j^{\perp} = \delta_{ij}$ that results in $\mathbf{b}_i^{\|} = \mathbf{a}_i^{\|}/2$ and $\mathbf{b}_i^{\perp} = \mathbf{a}_i^{\perp}/2$.

4D crystallography can be applied to obtain both aperiodic quasicrystalline and structurally similar periodic tilings, but but it does not offer any insight into physical mechanisms required for formation of quasicrystals. Therefore, we use the critical density wave approach, which goes back to the Landau idea that in a system near a phase transition, one can identify a critical degree of freedom, which determines the system's free energy. We combine this approach with an approximation of nonequilibrium assembly [20], which implies that the assembly is controlled by the binding energy between the growing cluster and a particle that can fill one of surrounding vacancies. Namely, at each step a set of positions that can be filled by a particle is determined, and the probability $p_j$ of filling the $j$-th position is calculated using the usual Boltzmann distribution

$$p_j = \exp\left(-\frac{E_j - \mu}{T}\right), \qquad (2)$$

where $E_j$ is the corresponding binding energy, $T$ is the thermal energy; the chemical potential $\mu$ is determined by the condition that exactly one particle is attached at each step. We also assume that the vacant positions are separated from the already occupied ones by one of the vectors $\mathbf{a}_i^{\|}$, where $i=0,1..8$. In Ref. 5 we applied a similar approach to the assembly of square-triangular dodecagonal clusters, but we expressed the binding energies not in the usual way (i.e., via interactions between particles in real space), but via 4D coordinates of the vacant positions. Here we show that the energies $E_j$ in Eq. (2) can be found without involving 4D crystallography, namely by using pair potentials obtained from the critical density wave $\rho(\mathbf{r})$, usually considered [21-24] in crystallization theory. The maxima of this function correspond to the positions of emerging structure and for ABT [Fig. 1(a)] it reads:

$$\rho(\mathbf{r}) = \sum_{i=0}^{3} \cos(2\pi \mathbf{B}_i^{\|} \mathbf{r} + \phi_i), \qquad (3)$$

where $\phi_i=0$, $\mathbf{B}_i^{\|} = \mathbf{b}_i^{\|} + \mathbf{b}_{i-1}^{\|} + \mathbf{b}_{i+1}^{\|} = \tau \mathbf{b}_i^{\|}$, and $\omega = \sqrt{2}+1$ is the coefficient of self-similarity. Function (3) is critical because the values of $\left(B_i^{\|}\right)^{-1}$ are close to the minimum distance between the ABT positions, and structural amplitudes $A(\mathbf{B}_i^{\|})$ are high; for more details see Supplemental Material, subsection A [25].

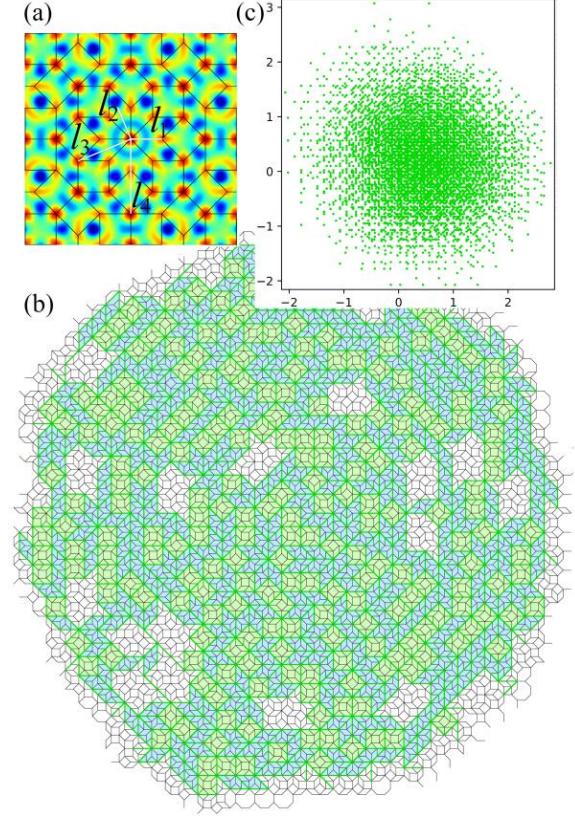

FIG. 1. Non-equilibrium assembly of random octagonal clusters. (a) Ammann-Beenker tiling superimposed on the critical density wave. Colors from red to violet correspond to the function values from 4 to -4. (b) Example of a cluster of 5000 particles (T=2). Perfection of the structure is justified by low defectivity of the second order tiling (light green squares and turquoise rhombuses that are $\omega$ times larger) and the small size of the cluster in perpendicular space. (c) Perpendicular coordinates of the particles of the same cluster.

Below, we calculate the binding energy $E_j$ as

$$E_j(\mathbf{r}_j) = \sum_i V(\mathbf{r}_i, \mathbf{r}_j),$$

where $V(\mathbf{r}_i, \mathbf{r}_j)$ is a pair potential, and $i$ runs over the occupied positions. We consider two types of the pair potentials. First, we introduce the local wave potential $V(\mathbf{r}_i, \mathbf{r}_j) = \alpha\rho(l)$, where $l = \mathbf{r}_i - \mathbf{r}_j$ and $\alpha$ is a step function of certain distance $L_0$: if $l \leq L_0$ then $\alpha = -1$ else $\alpha = 0$. Since ordinary pair potential does not depend at all on the rotation of particle pair, the value of $\rho(l)$ is taken to be maximally symmetrical, i.e. with $\phi_i = 0$. Surprisingly, using this potential one can obtain random octagonal clusters that consist exclusively of square and rhombic tiles; see more detail in Subsection A of Supplemental Material [25].

Morphologically close clusters emerge when using a simpler discrete pair potential $V(l)$, which is non-zero only for four distances $l_n$ between particles: $V(l_n) = [-2.48; 5; -3.07; -3.71]$; the distances $l_n$ are shown in Fig. 1(a). In the potential, the negative (energetically favorable) interactions are calculated as $-\rho(\mathbf{l}_i)$. The large positive interaction ($V_2 \geq 5$) can completely eliminate defect tiles. Note that the prohibited interparticle distance $l_2 \approx 1.083$ corresponds to the perpendicular distance $l_2^\perp$ that separates the opposite vertices of the ABT acceptance domain. Unlike the maxima, the minima of $\rho(\mathbf{r})$ are indexed by half-integer indices, so the distance from the origin to the nearest deep minima equals $l_2$ only approximately.

When using the discrete potential at $T \approx 2$, the most perfect quasicrystalline tilings arise; see Fig. 1(b). Surprisingly, the resulting structures consist exclusively of square and rhombic tiles, despite being phason-disordered. In addition to the temperature deviation from $T = 2$, one can also assemble a less ordered tilings by excluding the $V_4$ interaction; for more details see Supplemental Material, subsection A [25].

Thus, we have considered how the pair potentials can be constructed from the critical density waves used in Landau theory. The aperiodic defect-free random tilings obtained in this Section may be of interest for a variety of applications, such as studying the localized states in photonic and phononic quasicrystals [26] or even analyzing possible Hamiltonian cycles in them [27]. In addition to modelling aperiodic octagonal tilings, the critical density wave approach can be useful for obtaining other various quasicrystalline structures.

### III. ANOMALOUS PROTEINACEOUS SHELLS AND THEIR MODELS

Once the flat dodecagonal or octagonal order has been reconstructed into a similar appropriate periodic structure, it can be mapped onto a sphere using a net of a suitable polyhedron. In the dodecagonal case, this can be either an icosahedron or a cube [5]. In the octagonal case, only a cube net is suitable for mapping, and the ABT must be rearranged into a structurally similar periodic approximant with a square lattice. This can be done using nD crystallography; see Supplemental Material, Section B [25]. However, the Landau theory presents a clearer way to obtain octagonal periodic tilings using the corresponding critical density functions $\rho'(\mathbf{r})$. The latter functions appear as a result of small appropriate shifts of the wave vectors in the function (3), and like in the quasicrystalline case, the function maxima determine the positions of tiling nodes. Since we consider the shells with cube nets, such a function $\rho'(\mathbf{r})$ must have the translational symmetry of the square lattice, the modified vectors $\mathbf{B}_i^{\parallel}$ (we denote them as $\mathbf{B}_i'$) must become vectors of its reciprocal space. Therefore, in this space the enumeration of possible approximants is reduced to the enumeration of almost regular octagons whose vertices are the nodes of the square lattice; see Appendix, Fig. 4(a). Analysis of phases $\phi_i$ in functions $\rho'(\mathbf{r})$ shows that if we do not distinguish between the enantiomorphic cases, each square periodic approximant has two different structures; see Appendix for more detail.

Approximants shown in Figs. 2(a)-2(c) are obtained in the above-described way. Also, they can be assembled using the local wave potential at $T \to 0$. To assemble the $\sigma$ phase [Fig. 2(a)], one can use the potential $V(l) = \alpha\rho'(l)$, where $\alpha_0 = 1.85$. For the assembly of the second and third approximant structures, a slightly more complex potential $V(l) = \alpha[\rho'(l) - 1]$, where $L_0$=2.5 and $L_0 = 3.3$, respectively, is suitable. We emphasize that, like in the quasicrystalline case, the function $\rho'(l)$ in the potential has the phases $\phi_i$=0, while the use of functions with nonzero phases does not result in regular structures. Therefore, only three tilings are obtained, namely those that correspond to the observed proteinaceous shells with octagonal local order; see Figs. 2(d-j).

The structures of the first two approximants consist of only two types of tiles. Therefore, the realization in nature of these structures and the shells derived from them seems to us the most probable: using the Protein Data Bank [28], we found numerous anomalous shells of 7 different structural types, which are derived from cubic nets cut from these approximants. Figures 2(a-c) show that the vertices of cubic nets can pass not only through equivalent, but also through non-equivalent 4-fold axes, which in both cases after gluing the nets become 3-fold axes of the shells. Therefore, since the nets are determined by net edge, the net indices can also be half-integer. Also note that cubic nets with edges of the same length can correspond to two different shells; see such an example in the bottom of Fig. 2(b).

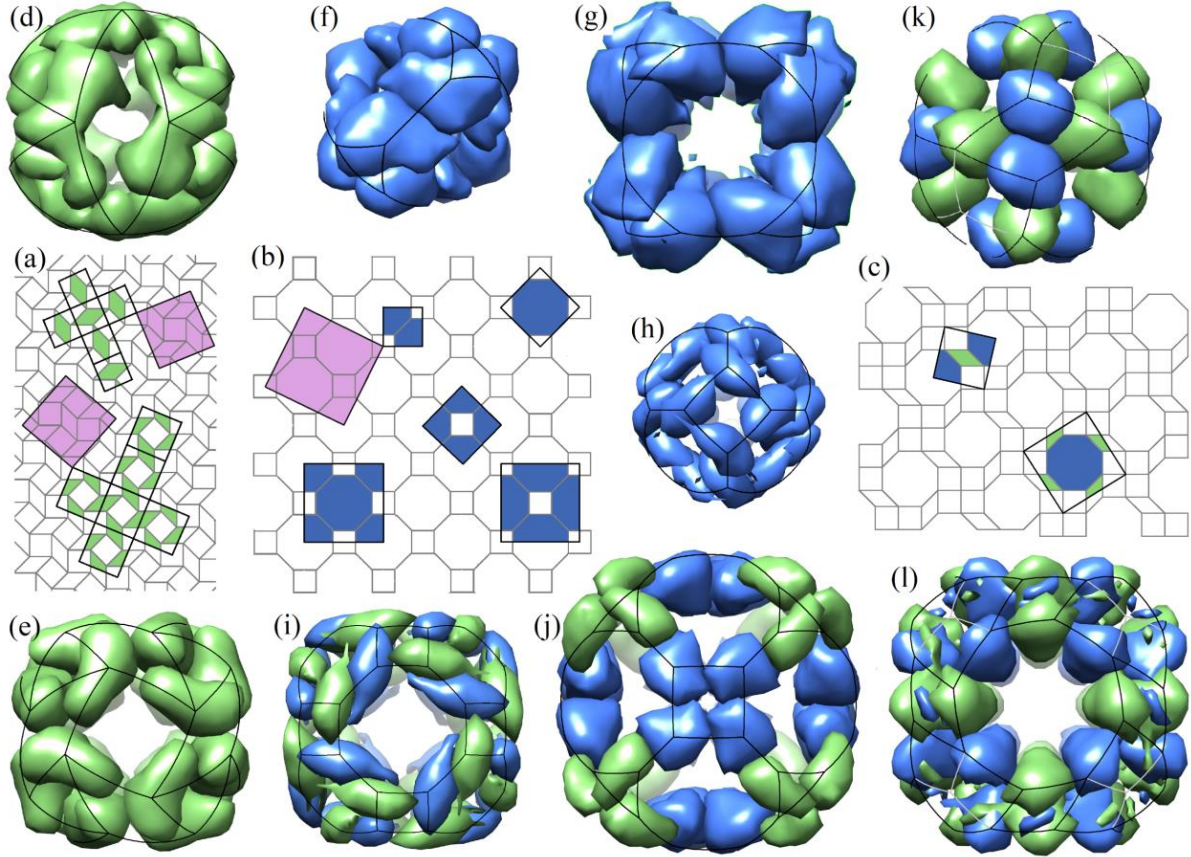

FIG. 2. The first three short-periodic octagonal approximants and spherical protein shells originated from them. (a) Nets for the shells demonstrated in (d) and (e) are colored in green and white. (b) Net faces for the shells shown in (f), (g), (h), (i), and (j) are in blue and white. (c) Net faces for the shells shown in (k) and (l). (d) 12-mer 3PV3. (e) 24-mer 6FDB. (f) 12-mer 3VAR. (g) 24-mer 2II4. (h) 24-mer 4I88. (i) 48-mer 4ELD. (j) 48-mer 6VFI. (k) 24-mer 2QAA. (l) 48-mer 8OPJ. Spherical quasi-lattices obtained as the result of energy minimization [see Eq. (4)] are superimposed on the observed shells.

When the nets cut from the third approximant [Fig. 2(c)] are glued, some nodes fall on cubic 3-fold axes. Asymmetric proteins cannot occupy such positions [29-30], so the latter remain unfilled. In spherical quasi-lattices (SQLs) superimposed on the observed structures the edges of square tiles containing unfilled positions are shown in gray [Figs. 2(k) and 2(l)]; in real shells, the bonds corresponding to these edges are absent.

The shape of the anomalous shells shown in Fig. 2 is more or less close to spherical. However, after gluing the cubic nets and mapping the tile vertices onto the sphere surface, the tiles become deformed, and the resulting SQLs must be optimized. Since the assembly of flat octagonal tilings is always controlled by interactions at two distances equal to the lengths of the tile edges and the diagonals of the rhombuses, respectively, the minimization of the following energy $E$ was used to optimize the SQLs:

$$E = \sum_{i>j}(|\mathbf{r}_i - \mathbf{r}_j| - r_0)^2 \\ + \sum_{k>l}(|\mathbf{r}_k - \mathbf{r}_l| - r_0')^2, \qquad (4)$$

where $\mathbf{r}_i$ are the coordinates of SQL vertices. The first sum in Eq. (4) runs over the SQL edges, and the second one runs over the longer diagonals of rhombuses. The values $r_0$ and $r_0'$ are the corresponding average values. In planar case $r_0'/r_0 \approx 1.848$, while for spherical geometry this ratio becomes smaller since the rhombuses are bent. Even if there are no rhombuses in the shell, minimization of the energy Eq. (4) equalizes distances of the same type, namely shorter diagonals of octagons. When rhombuses are absent, it is possible not to fix the ratio $r_0'/r_0$ and to minimize the energy over the variable $r_0'$. The arrangement of proteins in the considered shells is well described by SQLs obtained by minimizing energy Eq. (4) under a constraint that points $\mathbf{r}_i$ are retained at the spherical surface.

Our analysis of quasicrystalline proteinaceous shells presented above shows that the bonds between proteins in these assemblies almost always correspond to the edges of SQLs. Additional bonds (along the short diagonal of rhombuses) are present only in 6FDB and 2QAA shells. We also note that the number of symmetrically non-equivalent bonds in the shells is small (not more than 3), which is consistent with the CK quasi-equivalence principle, according to which proteins tend to organize equivalent, or if it is impossible, at least quasi-equivalent bonds with their neighbors. In the case of octagonal shells this principle leads to the formation not only of hexamers, but also of 4-fold and 8-fold capsomers, which are prohibited in original CK model [3]. Along with the intrinsic curvature of proteins, the formation of certain angles between interprotein bonds is also important for the assembly of such shells. Since there are no real bonds between proteins separated by a distance $\sim r'_0$, the second term in Eq. (4), being effective, leads to the appearance of angles ~135° between the appropriate pairs of bonds.

## IV. PREDICTED SHELLS: FACETING AND DISTRIBUTION OF TARGET TOPOLOGICAL CHARGES

As we have already noted, the experimental realization of shells based on the approximants shown in Figs. 2(a) and 2(b) is more probable, therefore in Fig. 3 (upper row) we present three spherical model tilings derived from these approximants and corresponding to yet undiscovered shells. More predicted structures can be found in Supplemental Material, see Section C [25].

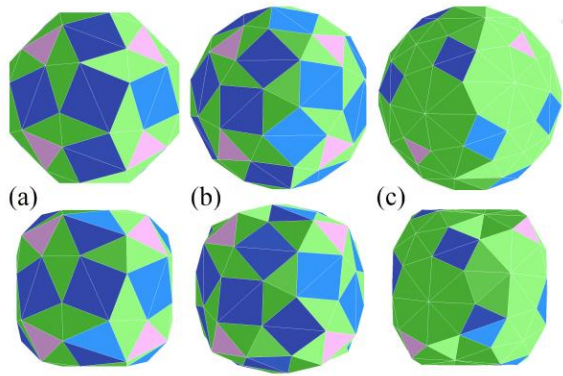

FIG. 3. Models of yet undiscovered shells containing 48, 72, and 60 proteins. The faces of corresponding nets are shown in Figs. 2(a) and 2(b) in pink. The shells in the second row differ from those in the first one by having a cubic faceting. In the shell (c), hexagons and octagons (representing hexamers and 8-fold capsomers) are decorated with triangles for better visualization only.

If we minimize the energy Eq. (4) of the model shells without additional constraints, then a cubic faceting can arise; see the bottom row in Fig. 3. However, the faceting does not manifest itself sharply, and below we explain this feature in terms of the distribution of topological charges over the shell vertices.

Recall that the topological charge of a polyhedron vertex is the difference between $2\pi$ and the sum of the values of all angles entering this vertex. For a convex polyhedron, the total topological charge always equals $4\pi$. Any polyhedron that is topologically equivalent to a sphere has the same topological charge, but, in contrast, the charge of a torus triangulation is zero.

We introduce a target topological charge and define it analogously, but using the angle values of the undistorted tiles. The vertex with nonzero target charge can obtain the real topological charge (and become a source of Gaussian curvature in the shell) provided the tiles entering the vertex take or tend to take their initial undistorted shape. It is easy to verify that in the anomalous shells shown in Figs. 2(k-i), and 2(l) all vertices have non-zero target topological charges, which are more or less uniformly distributed between them. This correlates well with the fact that the mass centers of proteins in these shells are located on near the spherical surface. The small visible faceting of the assemblies is mainly associated with their cubic symmetries and anisotropy of protein molecules.

In the anomalous shell presented in Fig. 2(j) protein positions shown in blue have zero target charges and the cubic faceting of the shell is more pronounced: the radii of protein mass centers for the proteins shown in blue and green are $r_1 \approx 7.17$ and $r_2 \approx 9.19$ nm, respectively. Minimization of energy Eq. (4) without constraints yields a smaller faceting, which can probably be explained by the fact that the proteins in the structure 6VFI are nonequivalent and octagons are deformed.

If all the vertices of the shell are symmetrically equivalent, then they obviously have equal charges. Surprisingly, the target topological charges are also equal in the nearly spherical shell shown in Fig. 2(i). In contrast, in the shells with three or more nonequivalent positions, uncharged vertices should appear, and the cubic faceting, depending on the elastic properties of the shell, can become more pronounced.

Note, however, that anomalous shells violating the CK model are limited in size: the largest anomalous shell known to us contain 72 pentamers [8]. Larger shells always satisfy the CK model, which provides a higher degree of protein quasi-equivalence. In the considered small shells the CK quasi-equivalence principle still continues to work and the formation of symmetrical shells consisting of regular tiles allows proteins to minimize the number of their environments.

## V. CONCLUSIONS AND OUTLOOK

Overall, we have shown that there exists a wide range of protein shells possessing octagonal quasicrystalline local order and predicted the structures of yet undiscovered analogous shells. In this context, we note that almost half of all known symmetrical enzyme complexes [28], the total number of which exceeds a thousand, satisfy the structural models shown in Figs. 2(d)-2(i) and 2(k). The quasicrystalline nature of the considered shells manifests itself both in the arrangement of individual proteins and interprotein bonds, which is especially important in the development of technologies for the controlled growth of protein nanoassemblies for specific applications. Compared to natural protein nanoassemblies, artificial nanoparticles exhibit a wider spectrum of anomalous structures [15-17, 31-33], many of which satisfy the proposed models; see Figs. 2(e), 2(f), 2(j) and 2(l). Our method of modeling the octagonal structures is based on Landau theory of critical density waves [21-22] and can be easily generalized for other types of quasicrystalline symmetry, which increases the predictive power of our approach. In addition, the results obtained are useful for further applications of the identified shells, studies of their self-assembly, thermodynamics, and mechanics including the mechanism of possible cubic faceting.

## ACKNOWLEDGMENTS

S. R., A.R., and O. K. acknowledge financial support from the Russian Science Foundation, Grant No. 22-12-00105-П. It is with great sadness that we announce that our co-author Rudolf Podgornik passed away on December 28, 2024.

## APPENDIX. STRUCTURES OF PERIODIC APPROXIMANTS FROM CRITICAL DENSITY FUNCTIONS

The reciprocal space vectors $\mathbf{B}'_i$ appearing in the periodic density functions $\rho'(\mathbf{r})$ coincide with the vertices of colored octagons superimposed on the square lattice; see Fig. 4(a). In the three short-periodic approximants considered in this article, the vectors $\mathbf{B}'_0$ and $\mathbf{B}'_1$ coincide, respectively, with the following ones: (<2,1>, <1,2>), (<3,0>, <2,2>), (<3,2>, <4,-1>). The other two wave vectors in $\rho'(\mathbf{r})$ are obtained by 4-fold rotation. Note that the third and fifth octagons, shown by dashed lines, after the rotation by 45° and certain homogeneous compression coincide with the first and second ones, respectively, so they result in the same tilings.

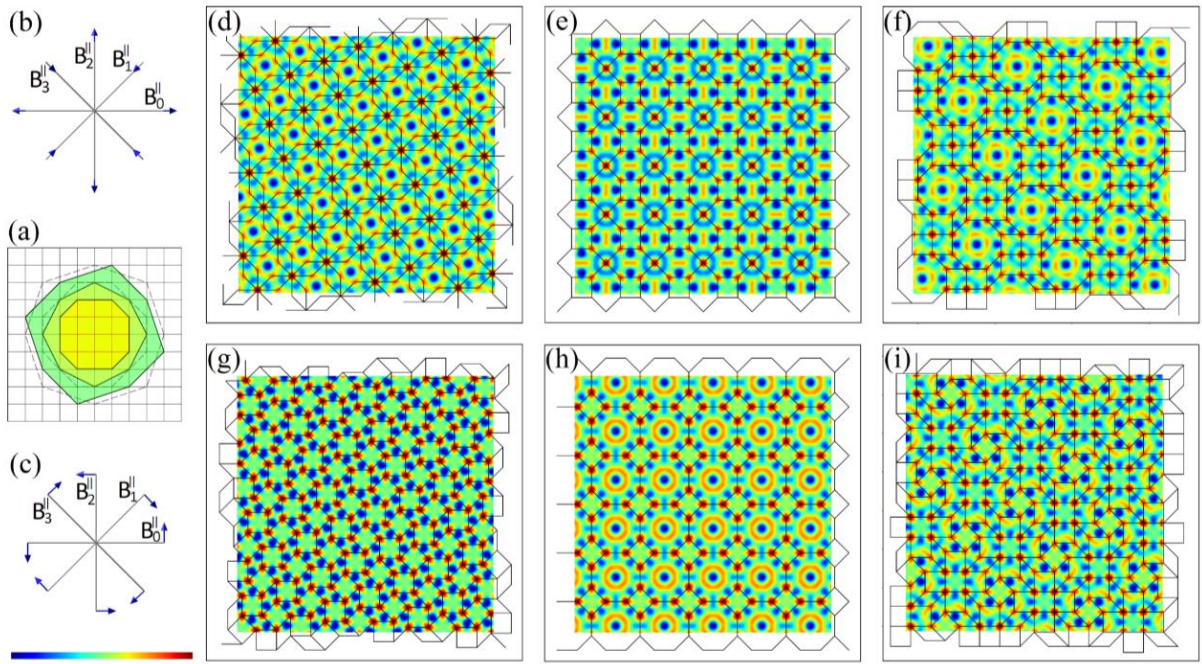

FIG. 4. Structures of the first short-period approximants of Ammann-Beenker tiling. (a) Nearly regular octagons whose vertices coincide with the nodes of the square lattice. (b)-(c) Irreducible shifts of the vectors $\mathbf{B}^{\|}_i$ leading to the appearance of commensurability between them. These shifts are spanned by two different one-dimensional irreducible representations of the $C_{8v}$ group and realized in the approximants presented in the first and second columns of the figure, respectively. In the third approximant, the shifts are reducible. (d)-(i) The first short-periodic approximants and their density functions $\rho'(\mathbf{r})$. Structures with the same periodicity are placed one above the other. The change in the background color from violet to red corresponds to the change in the value of the density function $\rho'(\mathbf{r})$ from minimum (-4) to maximum (+4).

Note that the presence of 4-fold translational symmetry imposes strict restrictions on the phases $\phi_i$ possible in $\rho'(\mathbf{r})$. In addition to the cases where all phases are 0 or $\pi$, 4-fold axis is preserved when the phases of adjacent plane waves alternate at the origin. However, at shift $<a/2, a/2>$, where $a$ is the lattice period, the 4-fold axis is superimposed with non-equivalent 4-fold one, and the phases in $\rho'(\mathbf{r})$ are changed, remaining still equal to 0 or $\pi$, which reduces the number of different density functions with the same periodicity. In the structures presented in the upper row of Fig. 4, the phases of all plane waves at the coordinate origin are zero. In the bottom row, alternative approximant structures with the same periodicity are presented. Also, we note that the approximants in Figs. 3(f), 3(g) and 3(i) are chiral, so the change of sign of the function $\rho'(\mathbf{r})$ results in the structure with an opposite handedness.

The vertices of all six tilings constructed correspond to the positions of the maxima of the six density functions found. In the tilings in Figs. 4(e) and 4(f), the positions at the least intensive maxima are excluded; filling them in the tiling (e) would result in too short distances between particles.

The periodic tilings shown in Figs. 4(g), 4(h) and 4(f) can be assembled within the approach of the local wave potential described in the main text. In the functions $\rho'(\mathbf{r})$ determining this potential $\phi_i = 0$. Exactly such functions are shown in the top row of Fig. 4. It is surprising that the potentials based on the functions shown in Figs. 4(d) and 4(e) lead to the assembly of tilings shown in g and h, respectively. Handedness of the chiral packings [Figs. 4(g) and 4(f)] emerges spontaneously during their assembly. If one excludes the linear term in the potential during the assembly of the tiling shown in Fig. 4(f), the positions inside the octagons will be filled.

We note that although in the framework of the local wave potential clusters with the structures shown in Figs. 4(d), 4(e) and 4(i) are not assembled starting from a single particle, they will continue to grow around a preassembled nucleus with a size of the order of $L_0$.

Also, we have considered the use of discrete potentials instead of the local wave ones. The approximant shown in Fig. 4(g) is assembled in about half of the cases using a simple potential with the binding energies $V_i = (-2.25, -4)$ at distances $l_1$ and $l_3$; see these distances in Fig. 1(a), the values of $V_i$ are found as $\rho'(l_i)$. In the second half of the cases, where the initial randomly formed cluster differs from the $\sigma$ phase nucleus, a locally defect-free aperiodic packing grows. The second approximant [Fig. 4(h)] is assembled using the potential $V_i = (-1, 1, -1, -1)$ that is non-zero at distances $l_i \approx (1, 1.732, 1.848, 2.414)$; here $l_2 = \sqrt{3}$, see other $l_i$ in Fig. 1(a). We did not find a simple discrete potential for the assembly of the tiling shown in Fig. 4(f).

# Supplemental Material for

## «Anomalous Proteinaceous Shells with Octagonal Local Order»

Sergei B. Rochal[1,*], Aleksey S. Roshal[1], Olga V. Konevtsova[1] and Rudolf Podgornik[2,3]

[1]*Faculty of Physics, Southern Federal University, 344090 Rostov-on-Don, Russia*

[2] *School of Physical Sciences and Kavli Institute for Theoretical Sciences, University of Chinese Academy of Sciences, 100049 Beijing, China*

[3] *CAS Key Laboratory of Soft Matter Physics, Institute of Physics, Chinese Academy of Sciences, 100190 Beijing, China*

**A. Nonequilibrium assembly of octagonal clusters**

To find the critical density wave one can use the ABT diffractogram. Diffraction intensity is obtained as $I(\mathbf{q}) = A(\mathbf{q})A^*(\mathbf{q})$, where q is the reciprocal space vector. Structural amplitude $A(\mathbf{q})$ reads:

$$A(\mathbf{q}) = \sum_n \exp(i\mathbf{q}\mathbf{r}_n),$$

where the summation is performed over all particles. Let us note that intensity of the central point of the diffraction pattern is $I(0) = N^2$. In the case shown in Fig. 1(a), $N = 5000$ and $I(0) = 2.5 * 10^7$. To make the figure clearer we used shades of blue for intensities between 0 and $2 * 10^5$, whereas all regions with higher intensities were colored in plain white. The red segment in Fig. S1(a) is equal to the inverse length of the short diagonal of the rhombus, is directed in the same way as the vector $\mathbf{B}_2^{\parallel}$, and is only slightly longer than it. The value of $|A(\mathbf{B}_i^{\parallel})|$ is close to 0.7 of the amplitude of the central peak, whereas the amplitudes of waves with shorter wave vectors are smaller by at least an order of magnitude. Due to these reasons, the maxima of critical density wave (3) correspond well to the ABT positions.

Usually, in models of nonequilibrium growth of a cluster, the vacant positions for filling are determined as minima of the binding energy [1-2]. Let us consider how this assumption differs from the assumption used in this article that the vacant positions are separated from already occupied positions by one of the vectors $\mathbf{a}_i^{\parallel}$, where $i$=0,1..8. Fig S1(b) shows a small octagonal cluster assembled within the latter approach at $T$=0.5 and $L_0 = l_4$; see distances $l_i$ in Fig. 1(a). Figure S1(b) shows that all local minima of the binding energy located near the cluster are separated from already filled positions by vectors $\mathbf{a}_i^{\parallel}$, although not all vacant positions, which are separated from the cluster by these vectors, correspond to local minima. However, filling such positions is energetically disadvantageous and therefore unlikely. Accordingly, when calculating the filling probabilities, the check that a given position corresponds to a local minimum can be omitted.

Figures S1(c) and S1(d) show two examples of clusters assembled at different values of temperature $T$ and parameter $L_0$. When $L_0 = l_3$, the local wave potential contains two coordination circles of minima only; more structurally perfect clusters arise if the next coordination sphere is included. Also, we note that the cluster (c) assembled at $T$=0.5 contains fragments with one-dimensional periodicity. With an increase in temperature, the structure becomes more homogeneous [see Fig. S1(d)], but single point defects appear in it.

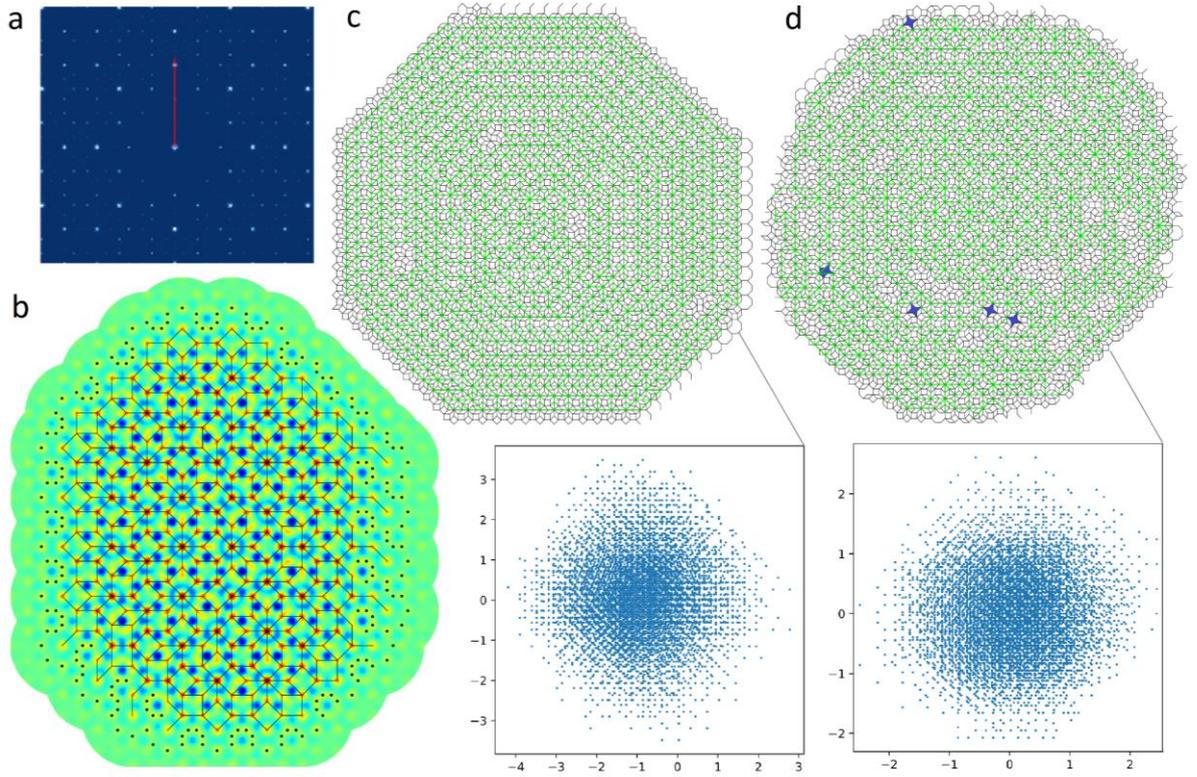

FIG. S1. Nonequilibrium assembly of octagonal clusters. (a) Diffraction from Ammann-Beenker tiling. (b) Field created by a cluster of 300 particles assembled in the local wave potential approximation with $L_0 = l_4$. The color change from red to blue corresponds to the change in the binding energy value from negative to positive. The black circles show the vacant positions considered within our approach. (c) Example of nonequilibrium assembly of 5000 particles ($T$=0.5) into a cluster using the local wave potential with $L_0 = l_3$. At $T \leq 0.5$, point defects never appear and the clusters consist exclusively of square and rhombic tiles. (d) A cluster of 5000 particles assembled at $T$=2 and $L_0 = l_4$. Single point defects are highlighted. Structural perfectness of the clusters can be estimated from the degree of defectivity of the 2nd order tiling, whose edge length is equal to $\omega$ (green lines) and from the cluster size in perpendicular space; see the corresponding cluster projections at the bottom of the figure.

**B. Periodic approximants in 4D crystallography**

In the conventional theory of QCs, linear approximants of the tiling can be obtained due to a certain linear phason strain [3], which changes only the perpendicular projections of the 4D nodes. Then, the modified perpendicular projections $\left(\mathbf{r}_j^\perp\right)'$ of j-th node read

$$\left(\mathbf{r}_j^\perp\right)' = \mathbf{r}_j^\perp + \hat{\boldsymbol{\varepsilon}} \mathbf{r}_j^{||}, \tag{S1}$$

where $\mathbf{r}_j^{||}$ are its parallel coordinates, and $\hat{\boldsymbol{\varepsilon}}$ is a tensor of the linear phason strain. A node is included in the tiling if its perpendicular projection belongs to the acceptance domain (AD). For ABT the AD shape is defined as the perpendicular projection of one cell of the 4D lattice, and in fact, the distances between opposite vertices of AD are perpendicular projections of vectors symmetrically equivalent to the vector <1,-1,1,1>. Under a phason strain AD is also deformed and its vertices are shifted according to eqn (S1). Placing the AD center at the coordinate origin results in the approximant structures shown in the top row of Fig 4. With suitable shifts of AD as a whole, the other approximant structures can also be obtained, but examining these shifts is much more difficult than considering possible critical density functions. Finishing the discussion of issues related to AD shape we recall that the length $l_2 \approx 1.083$ of the parallel projection of the same vector <1,-1,1,1> appears in the main text as the distance at which the placement of particles in random octagonal tilings is energetically unfavorable.

The irreducible linear phason strain tensors that reduce the symmetry to $C_{4v}$ have the following form [3]:

$$\begin{pmatrix} 0 & -\alpha \\ \alpha & 0 \end{pmatrix} \text{ and } \begin{pmatrix} -\beta & 0 \\ 0 & -\beta \end{pmatrix}. \tag{S2}$$

Superposition of both strains reduces the symmetry to $C_4$. The explicit form of the tensor for a given approximant is defined by its periodicity. Let this periodicity be characterized by two 4D translations $\mathbf{A}_1$ and $\mathbf{A}_2$. Then, the components of the linear phason strain tensor $\hat{\varepsilon}$ are expressed through standard relations $\mathbf{A}_1^\perp = \hat{\varepsilon}\mathbf{A}_1^\parallel$ and $\mathbf{A}_2^\perp = \hat{\varepsilon}\mathbf{A}_2^\parallel$.

Using the vectors $\mathbf{a}_i^\parallel$ and $\mathbf{a}_i^\perp$ and (S1), one can find the vectors $\mathbf{a}'^\perp_i$ in the approximants. Then taking into account that $\mathbf{b}'_i\mathbf{a}'_i = \delta_{ij}$ one can express the vectors $\mathbf{b}'^\parallel_i$ as $\mathbf{b}'^\parallel_i = \mathbf{b}^\parallel_j - \hat{\varepsilon}^T\mathbf{b}^\perp_j$. Accordingly, the vectors $\mathbf{B}'_i \equiv \mathbf{B}'^\parallel_i$ that determine the explicit form of density function $\rho'(\mathbf{r})$ take the analogous form:

$$\mathbf{B}'_i = \mathbf{B}^\parallel_j - \hat{\varepsilon}^T\mathbf{B}^\perp_j,$$

where $\mathbf{B}^\parallel_i = \omega\mathbf{b}^\parallel_i$, $\mathbf{B}^\perp_i = -\omega^{-1}\mathbf{b}^\perp_i$. The shift of the vectors $\mathbf{B}^\parallel_j$ under the phason strains is shown in Figs. 4(b) and 4(c), respectively.

In this paper, we consider three shortest-periodic square-lattice approximants for which $|\mathbf{A}_i^\perp|/|\mathbf{A}_i^\parallel| \le \tau^{-1} = \tau - 1$. The first three structures satisfying this condition are characterized by the following periodicities $\sqrt{2+\sqrt{2}} \approx 1.848$, $1+\sqrt{2} \approx 2.414$, and $\sqrt{6+3\sqrt{2}} \approx 3.200$. In the corresponding phason strain tensors the coefficients α and β read ($\alpha = -\omega^{-1}$, $\beta=0$), ($\alpha = 0$, $\beta=\omega^{-2}$), and ($\alpha = \omega/3$, $\beta=-2\sqrt{2}\omega^{-1}/3$), respectively.

## C. Four more structures predicted and some general remarks

We do not consider the spherical tilings derived from two approximants shown in Figs. 4(d) and 4(e) because these tilings differ significantly from the packings of observed protein shells with octagonal local order. Such shells known to date have at most $N$=48 protein positions. We restrict our consideration to model tilings that have at most $N$=72 positions not located on the axes of symmetry. In addition to Fig. 3, Fig. S2 shows 4 more models of yet undiscovered shells with octagonal local order. These models are derived from two variants of the approximant structures shown in Figs. 4(i) and 4(f). Regularization of tiles in the shells was carried out according to the procedure described in the main text, however, an additional term $\sum_{m>n}(|\mathbf{r}_m - \mathbf{r}_n| - r_0'')^2$, in which the summation was carried out over the diagonals of the square tiles, was added to energy (4). In this term, the value of $r_0''$ is the average length of the diagonals. Note that to construct the model shells shown in Fig. 3(c), an analogous term was added to the energy and the summation was performed over the diagonals of hexagonal tiles lying around the 3-fold axes.

Note that if a cubic net of some shell does not contain positions at the vertices of its faces, then the resulting shell has $N = 6m(h^2 + k^2)$ positions, where $m$ is the number of positions per primitive cell of the approximant, while $h$ and $k$ are indices of the cubic net cut from its structure. Since the indices of the cubic nets are either both integer or both half-integer [4], the term $6(h^2 + k^2)$ takes the following values: 3, 6, 12, 15... If the position lies at the vertex of the net faces, then the calculation of the position number must take into account the exclusion of eight $90^0$ sectors when the net is glued together. Because of this exclusion, for shells with 4 and 8 vertices on the 3-fold axes, $N$ must be increased by 1 and 2, respectively.

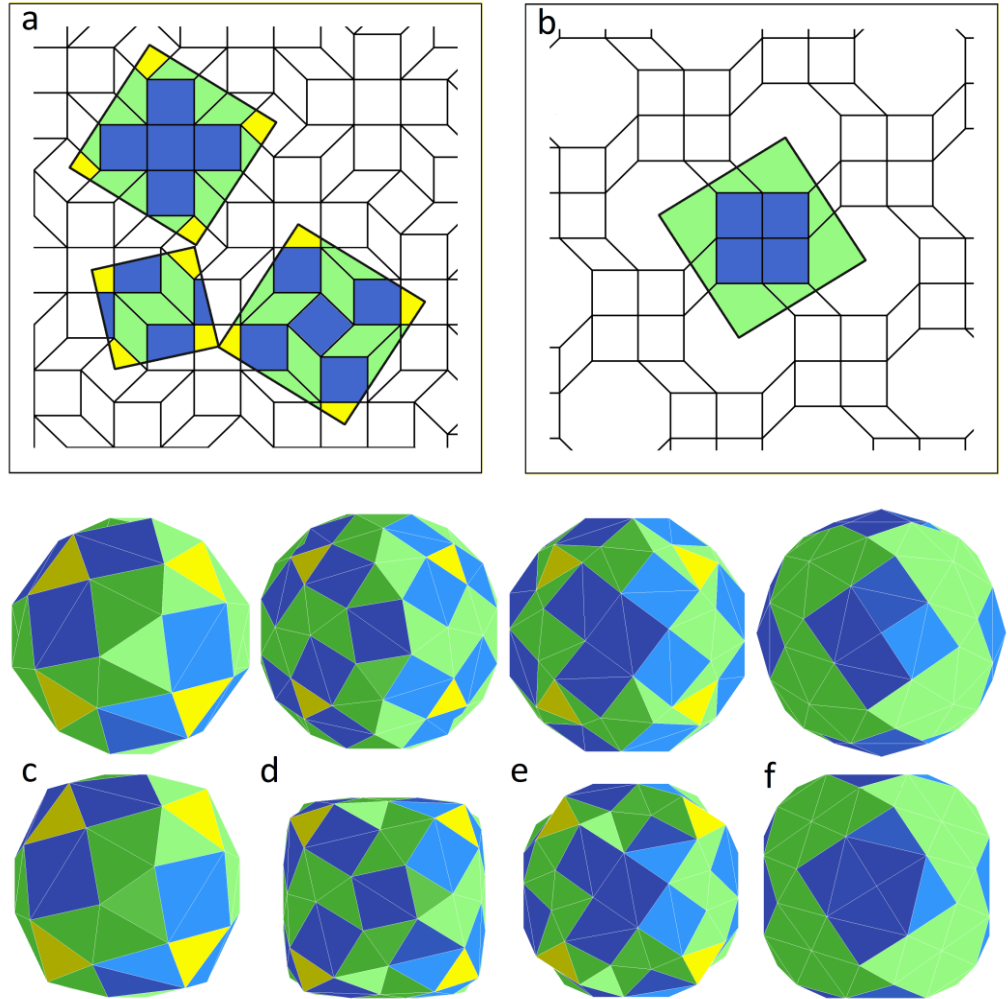

FIG. S2. Square nets of yet undiscovered octagonal shells and their models. (a)-(b) Approximant structures and faces of the cubic nets under consideration. (c)-(f) Resulting spherical tilings. The upper row shows spherical shells, while the bottom row shows the same shells with cubic faceting. The first three shells are obtained from the approximants demonstrated in (a) and have 36 (T symmetry), 72 (O symmetry), and 72 (O symmetry) protein positions, respectively. The last shell is derived from the approximant in (b) and has O symmetry. Of the 62 vertices that the shown polyhedron has, 8 vertices on the 3-fold axes are added to visualize the shell better. Since proteins cannot be located on the symmetry axes, a possible protein assembly corresponding to the polyhedron can contain only 48 proteins; see similar position exclusion in the 8OPJ shell discussed in the main text.